# Atomic gravimeter robust to environmental effects

**Authors:** Cristian D. Panda*[1], Matt Tao[1], Miguel Ceja[1], Andrew Reynoso[1], Holger Müller*[1]

[1]Department of Physics, University of California, Berkeley, 94720, CA, USA.

*Corresponding authors. Email: cpanda@berkeley.edu, hm@berkeley.edu

**Abstract:** Atomic accelerometers and gravimeters are usually based on freely-falling atoms in atomic fountains, which not only limits their size, but also their robustness to environmental factors such as tilts, magnetic fields or vibrations. Such limitations have precluded their broad adoption in the field, for geophysics, geology, and inertial navigation. More recently, atom interferometers based on holding atoms in an optical lattice have been developed. Such gravimeters also suppress the influence of vibrations in the frequency range of ~1 Hz and above by several orders of magnitude relative to conventional atomic gravimeters. Here, we show that such interferometers are robust to tilts of more than 8 mrad with respect to the vertical and can suppress the effect of even strong environmental magnetic fields and field gradients by using atoms in the $F = 3,4$ hyperfine ground states as co-magnetometers, potentially eliminating the need for shielding. We demonstrate gravimeter sensitivity of 0.7 mGal/$\sqrt{\text{Hz}}$ (1 mGal=10 µm/$s^2$) in a compact geometry where atoms only travel over mm of space.

**Introduction**

Atomic accelerometers and gravimeters have demonstrated performance rivaling classical sensors (*1–7*). They offer state-of-the-art accuracy and long-term stability, thanks to using the wavelength of a highly stabilized laser as a reference (*8, 9*). However, atomic sensors based on light-pulse atom interferometry with freely falling atoms in an atomic fountain (*10*) have seen limited adoption in the field. First, pulsed operation makes them excessively sensitive to vibrations due to aliasing. Second, the fountain principle requires vertical alignment to better than 1 mrad to avoid signal loss from atoms falling out of the laser beam. Third, magnetic-field control over the entire fountain height (which can be half meter or more) is necessary. Overcoming these limitations may require complex vibration isolation (*4, 11*), complex gimbals (*12*), and heavy and bulky shielding (*13, 14*).

Optical-lattice atom interferometers have shown unprecedented gains in coherence (*15–17*), being able to reach measurement times exceeding one minute (*18*). In addition, lattice-atom interferometers are orders of magnitude less sensitive to environmental vibrations than their atomic fountain counterparts (*17*). Here, we show that they can also address the above limitations. We demonstrate robustness to tilts as large as 8 mrad, as the optical lattice prevents the atoms from spilling out. We also show that they are robust against strong magnetic fields and field gradients. These cause only second order errors that, as we demonstrate, can be compensated for by using the $F = 3,4$ hyperfine states as co-magnetometers. Lastly, we demonstrate sensitivity improved by more than an order of magnitude compared to our previous publication (*17*), to $0.7$ mGal/$\sqrt{\text{Hz}}$. This level of precision is sufficient, e.g., for applications in inertial navigation with reference gravity maps (*2, 19*).

Atomic Interferometer Gravimeter

The basic operation of our lattice atom interferometry gravimeter has been described elsewhere (*18*). In brief, cesium (Cs) atoms are cooled to a motional temperature of 300 nK and transferred to the magnetically insensitive $m_F = 0$ state of the ground hyperfine manifold (Fig. 1(a)). A pair of Raman pulses act as a beamsplitter, creating a spatial superposition state where each atom is a superposition of two wavepackets (Fig. 1(b)). The wavepackets are loaded into a one-dimensional optical lattice formed by a far-off resonant, red-detuned laser ($\lambda = 943$ nm) coupled to the optical cavity. The wavepackets are separated by a vertical distance of $\Delta z = n\lambda/2$, where $n$ is an integer. The optical lattice is turned off after a time interval $\tau$ and Raman beamsplitters are used to recombine the two wavepackets.

The optical lattice potential

$$U_{\text{latt}} = -U_0 \exp\left(\frac{-(x^2 + y^2)}{2\,w_0^2}\right) \sin^2(\frac{2\pi z}{\lambda}) \qquad (1)$$

confines the atoms during the interferometer hold, where $U_0$ is the trap depth and $w_0 = 760$ $\mu$m is the cavity mode waist. The lattice confines the Cs atoms against maximum accelerations of $a_\parallel \simeq 2U_0/(\lambda\,m_{\text{Cs}})$ in the axial direction, where $m_{\text{Cs}}$ is the Cs atomic mass. The experiments described here were performed with $U_0 = 18\,E_r$, where $E_r = \frac{m_{\text{Cs}} v_r^2}{2} \simeq 2$ kHz $\cdot\,h$ is the Cs atom recoil energy when absorbing 852 nm photons and $h$ is the Plank constant. This holds the atoms against Earth's gravity, which typically only requires $U_0 > 5\,E_r$. The optical lattice also provides transverse confinement against accelerations up to $a_\perp \simeq U_0/(w_0\,m_{\text{Cs}})$. This is weaker than in the axial direction, $z$, since the cavity waist $w_0 \gg \lambda$, but still sufficient to hold atoms against experiment tilts of nearly 10 mrad relative to the vertical.

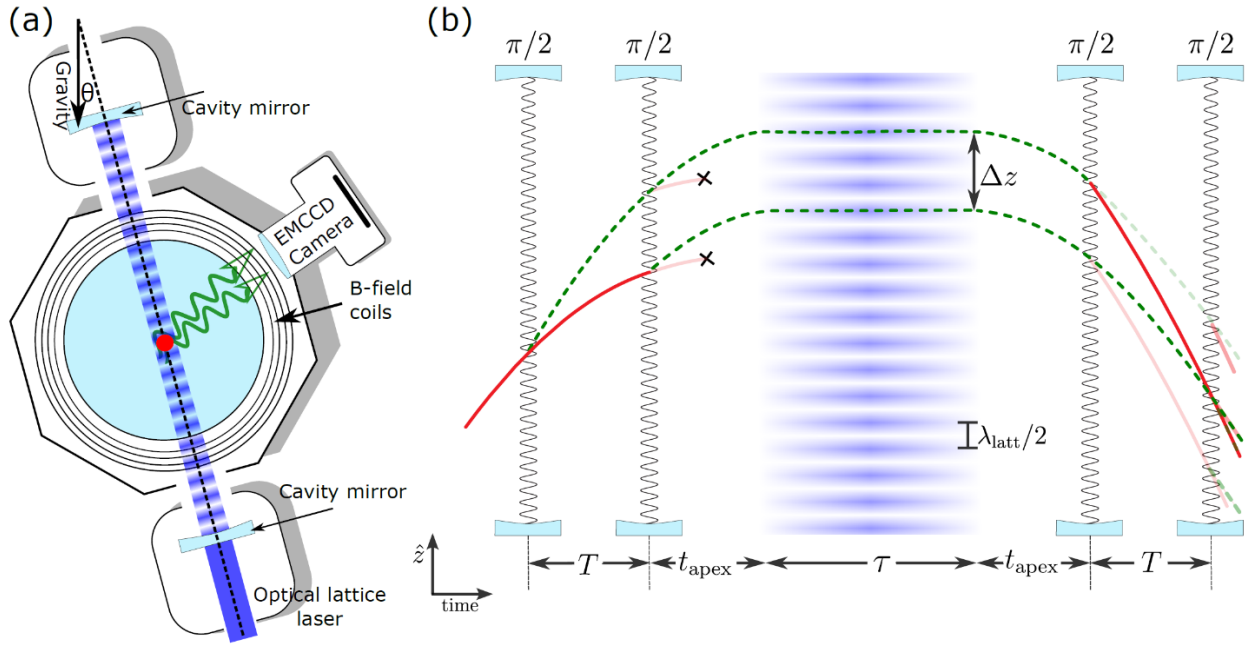

**Figure 1. Atomic gravimeter based on atom interferometry in an optical lattice.** (a) Cs atoms (red disk) in a spatial superposition are loaded into the high-intensity regions of an optical lattice, formed by a laser coupled into the Gaussian mode of an optical cavity. The measured phase is extracted from the ratio between the two interferometer ports, measured by collecting fluorescence photons (green wavy lines) on a camera. Adjusting the legs of the optical table tunes the tilt angle $\theta$ of the cavity. (b) Pairs of Raman beamsplitters create a superposition state of atomic wavepackets at two different axial positions, separated by $\Delta z$. The gravitational potential energy difference accumulated during the optical lattice hold represents a precise measurement of gravity.

The phase accumulated between the interferometer arms is dominated by the difference in the gravitational potential

$$\phi = \omega\,\tau = \frac{m_{Cs} g \Delta z \tau}{\hbar}. \tag{2}$$

The local gravitational acceleration $g$ can thus be extracted from the measurement of $\phi$.

The phase $\phi$ is read out by fluorescence detection as follows. A push beam separates the two interferometer ports and fluorescence from each port ($N_3$ and $N_4$) is collected by a lens and recorded on an EMCCD camera. We compute the asymmetry $A = \frac{N_3 - N_4}{N_3 + N_4}$, which is a sinusoidal function of the hold time $\tau$:

$$A(\tau) = C \cos(\phi - \tau \omega). \tag{3}$$

The fringe contrast, $C$, is a measure of the proportion of atoms participating in the interferometer and is at most 0.5 in this interferometer.

Gravity measurements with a tilted interferometer

The interferometer is mounted on an optical table, whose four legs have the usual pneumatic leveling and dampening system. We attach small electric motors (Futaba S148) to two of the four leveling valves, to control the tilt of the optical table with respect to Earth's gravitational axis. After each tilt change, we wait 5-20 minutes for the table position to settle. The experiment tilt angle, $\theta$, is independently measured along two axes using an electrolytic sensor tilt-meter (Applied Geomechanics 700) that is rigidly mounted to the interferometer setup. The tilt-meter has precision and accuracy of < 0.01 mrad.

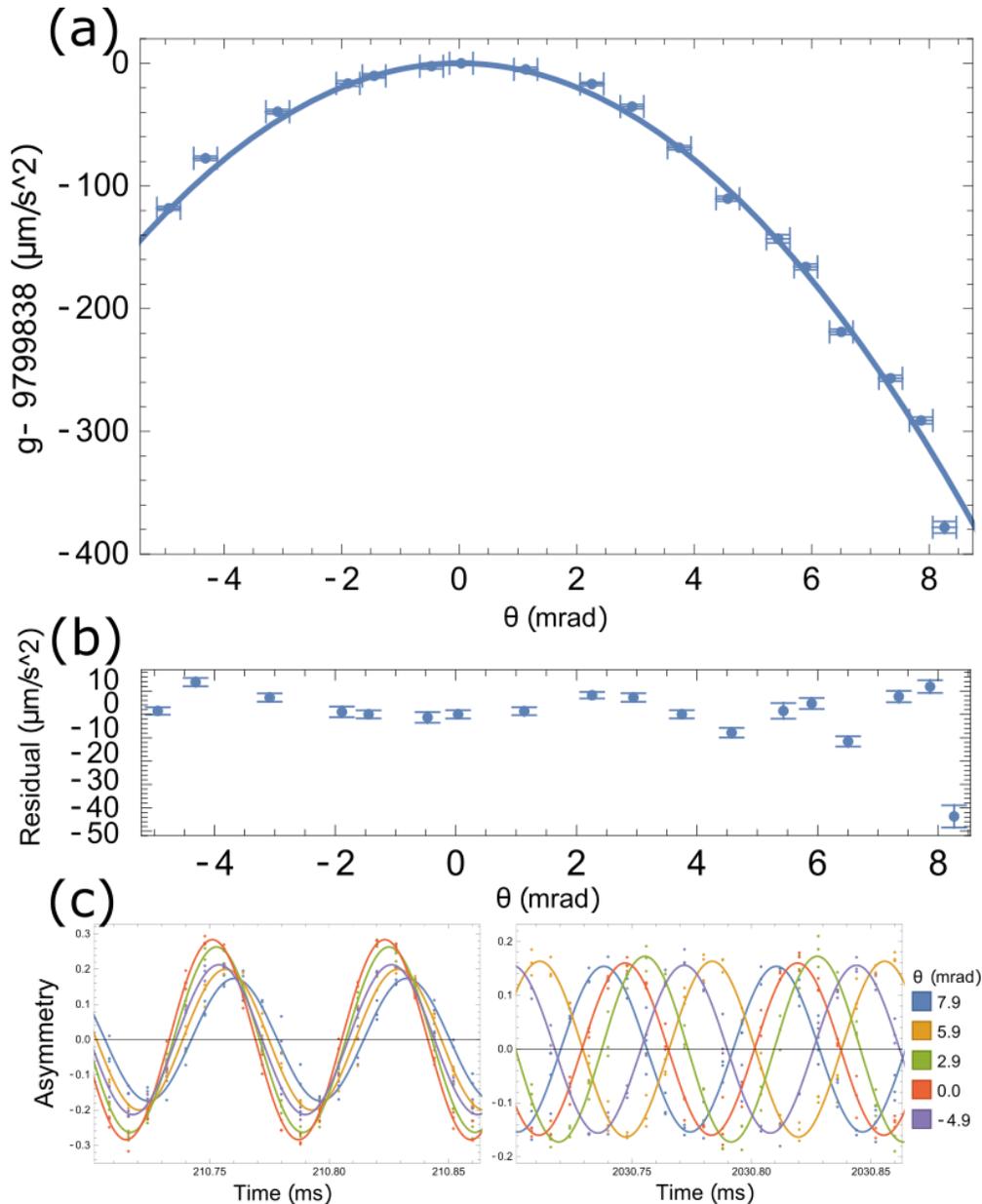

**Figure 2. Tilt measurement** (a) Measured value of laboratory gravity as a function of experiment tilt angle. Solid line corresponds to expected gravity value as a function of tilt. Error bars correspond to nonlinear fit $1\sigma$ errors. (b) Difference between measured and expected gravity values. The calculation of the residuals does not use the horizontal uncertainty in $\theta$. (c) Sample atom interferometric fringes at varying tilts for (left) short hold times and (right) long hold times. Large tilts lead to a smaller projection of Earth's gravity along the interferometer axis, therefore smaller fringe frequency and less phase for the same hold time.

During each gravity measurement, which takes 5-6 minutes, we observe typical tilt variation and drift of less than 0.1 mrad. Additional tilt variations at the 0.2 mrad level are due to variations in the alignment of the lower cavity mirror.

We measure gravity by recording fringes at both short hold times ($\tau_S$) and long hold times ($\tau_L$) (Fig. 2(c)). This separates out the phase terms accumulated during the free-fall part of the interferometer, which are not included in Eq. 2. This removes undesired systematic shifts and prevents frequency drifts from entering the lattice interferometry measurement.

For each gravitational measurement, long and short hold-time fringes are combined into one dataset. We fit this dataset to equation

$$A(\tau) = C_0 \cos(\tau\omega - \phi_0) \exp(-\tau/\tau_C) + d. \tag{4}$$

with free-parameters: the initial contrast $C_0$, the offset phase $\phi_0$, the fringe frequency $\omega$, the contrast decay constant $\tau_C$, and a parameter $d$ to account for fringe offsets due to background imaging scatter. $\tau_C$ accounts for the decrease in contrast at longer hold times.

The initial guess for $\omega$ must be correct to within $2\pi/\tau$ to avoid degeneracy of the fringe frequency and therefore remove any ambiguity in the measured $g$ value. To obtain this guess, we start with a coarse measurement with a low value of the hold time $\tau$ ($\tau \approx 210$ ms) and then record increasingly finer measurements of $g$, by increasing $\tau$ up to $\tau \approx 2030$ ms. Figure 3 shows all calibration measurement datasets and single fit to all datasets. The gravitational acceleration, $g$, is then extracted from the fitted fringe frequency, $\omega$, using Eq. 2. The $m_{Cs}/\hbar$ ratio is known to precision much beyond needed here (20, 21). The wavepacket separation $\Delta z = n\lambda/2$ is known from the laser wavelength, which we monitor with a calibrated wavemeter.

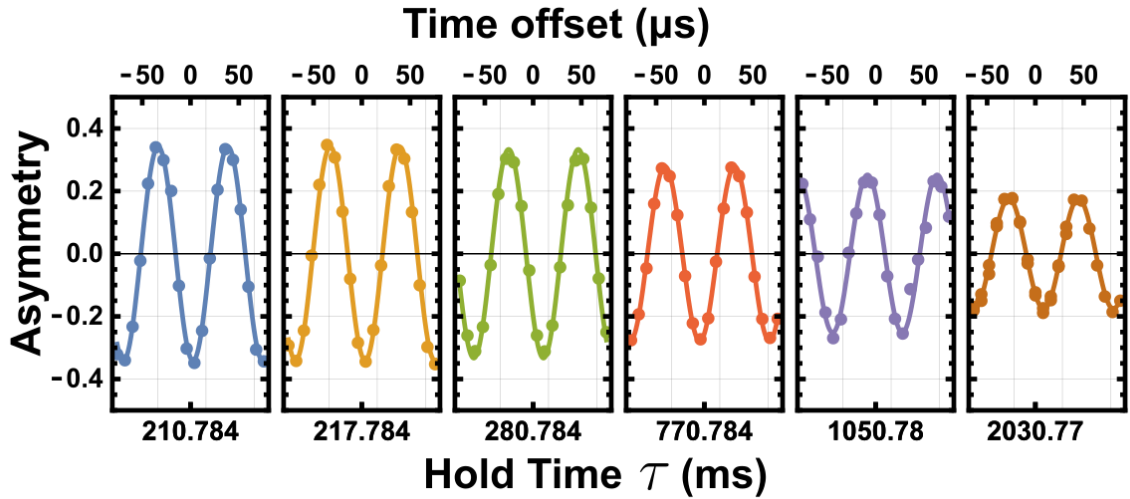

**Figure 3. Fringe number calibration measurement.** Data with varying hold times $\tau$ are iteratively fitted to extract the fringe frequency with much better than $2\pi/\tau$ accuracy, removing any possible ambiguity in the gravimetry measurement.

We then vary the interferometer tilt and measure the resulting local gravity values. The results are shown in Fig. 2(a). The solid line is the predicted change in the measured acceleration as a function of tilt, representing the projection of Earth's gravity along the axis of our experiment (see Fig. 2(a)),

$$g(\theta) = g_0 \cos(\theta). \tag{5}$$

We observe very good agreement between the experimentally measured $g$ and that expected for each tilt value within the error bars of the measurement. We note that the residuals computed in Figure 2(b) do not account for the large error bars in the measured tilt values.

This data demonstrates atom interferometry with more than 8 mrad of tilt between the interferometer and gravitational axes. We observe a factor of two loss in atom numbers and a small reduction in contrast, which we attribute to the fact that atoms spill out during the interferometer beam-splitter phase, when the atoms are in free-fall. This results in a factor of 2 loss in precision compared to zero tilt. In a gravimetry measurement, Eq. 5 is used to extract the vertical value of gravity, $g_0$, with corresponding uncertainty $\delta g_0$. This requires knowledge of $\theta$ to an accuracy beyond $(\delta g_0/g_0)/\theta$. For $\theta = 10$ mrad, achieving $\delta g_0/g_0 = 10^{-6}$, as needed for inertial navigation, requires measuring $\theta$ to an accuracy of 0.1 mrad, which our tilt sensor is capable off.

Operation at even larger tilts could be possible with increased lattice trap depth $U_0$, which increases radial confinement of the atoms. While this comes at the cost of reduced atom interferometer contrast, we show in a separate publication that reducing the atom sample temperature greatly boosts contrast (*18*). Another path to potentially achieve operation at larger tilts is using a separate blue-detuned laser beam coupled to the higher order Laguerre Gaussian 01 "donut" mode of the cavity. A far-off resonant blue-detuned laser with a trap depth of 50 recoils would provide sufficient transverse confinement to perform interferometry with tilt angles up to 20 degrees. This might open the door for interferometer operation on board of naval vessels without gimbals (*22*). Yet another approach could be to use two additional overlapped optical cavities to create a 3D lattice, providing equal confinement in all spatial dimensions (*23*). This geometry would make the interferometer agnostic to Earth's gravity axis, therefore achieving multi-axis inertial measurement of acceleration.

Response to large magnetic field offsets and gradients

To check the response of the lattice atom interferometer to environmental magnetic fields, we apply strong magnetic fields by turning on the MOT coils, which generate a magnetic field gradient of 15 G/cm and offset magnetic field of ~0.5-1 Gauss at the position of the atoms. We observe no loss in contrast at the 2% level. As expected, we observe a small phase shift accumulate during the interferometer hold (Fig. 4), consistent with that due to the second order Zeeman effect of the magnetic field gradient $\partial B_z/\partial z$ (*24*, *25*)

$$\phi^B = \frac{(g_J - g_I)^2 \mu_B^2}{2\hbar \, \Delta E_{\text{hfs}}} B \frac{\partial B_z}{\partial z} \Delta z \tau, \tag{5}$$

where $\mu_B$ is the Bohr magneton and $\Delta E_{\text{hfs}}$ is the Cs atom hyperfine splitting.

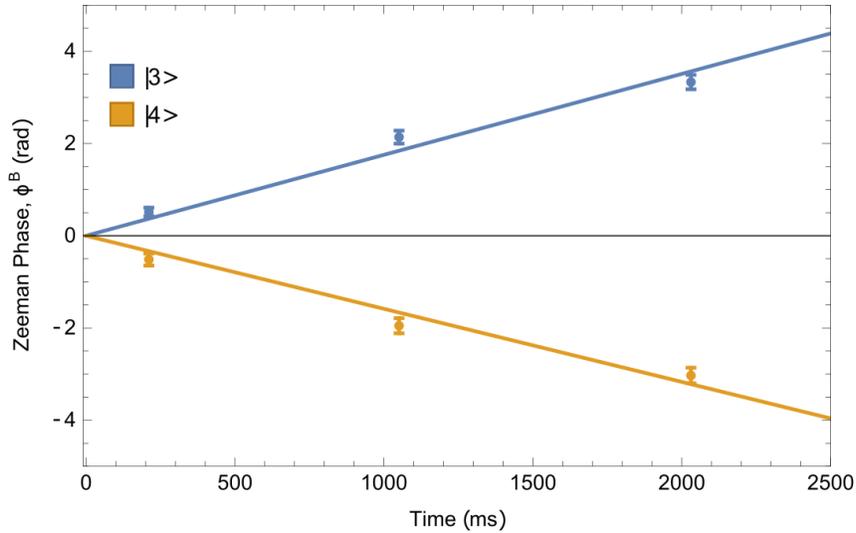

**Figure 4. Response to magnetic field gradients.** Phase shift observed by the interferometer in the presence of applied magnetic field gradients of 15 G/cm when atoms are in $|F = 3\rangle$(blue) and $|F = 4\rangle$(yellow). The magnitude of the shifts is consistent with the second order Zeeman effect in the Cesium atoms.

For typical environmental magnetic field gradients, which are 4-5 orders of magnitude lower than the field applied here (*26*), this systematic would cause a shift in the measured value of $g$ at the $10^{-7} - 10^{-8}$ m/s$^2$ level, which is negligible for most applications. The lattice interferometer could be used in the field without the need for bulky and costly magnetic shielding.

This phase shift reverses sign when performing the atom interferometer in either of the two hyperfine states $|F = 3\rangle$ and $|F = 4\rangle$ of the ground Cs electronic manifold. Taking the average of the two measurements means that this already small systematic effect can be further suppressed by a factor much beyond the factor of $\sim 50$ demonstrated here. This procedure effectively uses the two hyperfine states as a co-magnetometer.

Interferometer sensitivity

We run the experiment with a variety of parameters and choose the region of parameter space which optimizes experiment sensitivity. The data presented in this manuscript was acquired with interferometer separation of $\Delta z = 4.24$ $\mu$m, corresponding to 9 lattice sites and trap depth $U_0 = 18$ $E_r$. Each experimental shot measures $10^5 - 10^6$ atoms.

Upgrades in the experimental apparatus since our last publication (*17*) have increased the atom number 40-fold and decreased atom sample temperature by 40%. These improvements resulted in an experiment precision below 0.7 mGal/$\sqrt{\text{Hz}}$, more than an order of magnitude better than in our previous publication. The measurement uncertainty is consistent with the Standard Quantum Limit floor (*27*). We expect further gains in interferometer performance through further reduction in atom sample temperature. Even at current sensitivity, our interferometer would be compatible with inertial navigation in the field (*2*).

**Discussion**

We have described above a gravimeter based on an atom interferometer, where the atoms are held against gravity and other inertial forces by a one-dimensional optical lattice, formed by the Gaussian mode of an optical cavity. The precision of the atom interferometer is better than 0.7 mGal/$\sqrt{\text{Hz}}$, more than an order of magnitude better than in our previous publication (*17*). Operation near the standard quantum limit suggests that further upgrades in statistics or coherence are very likely to further increase precision.

The interferometer can perform gravimetry measurements robust to tilts beyond 8 mrad, larger than attainable with traditional, atom fountain light-pulse interferometers. This is possible because the optical lattice provides confinement along the axial direction of the interferometer and sufficient confinement in the transverse direction to prevent atom loss. In addition, the interferometer coherence is robust to environmental magnetic fields and gradients. A simple co-magnetometer scheme can cancel systematic shifts due to environmental magnetic field gradients. The experiment is performed in a compact setup, where the atoms only travel for 2 mm, compared to tens of cm or meters common in atomic fountains. Together with the known robustness of lattice-based interferometers against vibration (*17*), these significant advantages over atomic fountain light-pulse interferometers are a first step towards more robust sensors in the field, obviating the need for unwieldy magnetic shielding, complex vibration isolation schemes, and gimbals to correct for environmental tilts.

**Acknowledgments**

We acknowledge experimental contributions by James Egelhoff, and discussions with Matt Jaffee and Victoria Xu.

**Funding:** This material is based on work supported by the
National Science Foundation grants 1708160 and 2208029
Department of Defense Office of Naval Research grant N00014-20-1-2656
Jet Propulsion Laboratory (JPL) grants 1659506 and 1669913.

**Author contributions:** CDP and HM conceptualized the experiments. CDP built the apparatus, collected measurements, and analyzed data. MT, MC, AR and HM contributed to the building of the apparatus. CDP and HM wrote the original draft. CDP, MT, HM contributed to the review of the manuscript. HM supervised the experiment.

**Competing interests:** Authors declare that they have no competing interests.

**Data and materials availability:** All data presented in this paper is deposited online.